\documentclass[prd,twocolumn,preprintnumbers,amsmath,amssymb,superscriptaddress,nofootinbib]{revtex4}
\pdfoutput=1

\usepackage{epsfig,psfrag,graphics,verbatim}
\usepackage{graphicx}
\usepackage{epstopdf}
\usepackage{dcolumn}
\usepackage{bm}
\usepackage{slashed}
\usepackage{color}
\usepackage{amssymb}

\begin{document} 
\preprint{DESY 12-174}
\preprint{TUM-HEP 866/12}

\title{On the spin-dependent sensitivity of XENON100}

\author{Mathias Garny}
\affiliation{Deutsches Elektronen-Synchrotron DESY, Notkestra\ss{}e 85, 22603 Hamburg, Germany}

\author{Alejandro Ibarra}
\affiliation{Physik-Department T30d, Technische Universit\"at M\"unchen, James-Franck-Stra\ss{}e, 85748 Garching, Germany}

\author{Miguel Pato}
\affiliation{Physik-Department T30d, Technische Universit\"at M\"unchen, James-Franck-Stra\ss{}e, 85748 Garching, Germany}

\author{Stefan Vogl}
\affiliation{Physik-Department T30d, Technische Universit\"at M\"unchen, James-Franck-Stra\ss{}e, 85748 Garching, Germany}

\date{\today}

\begin{abstract}
The latest XENON100 data severely constrains dark matter elastic scattering off nuclei, leading to impressive upper limits on the spin-independent cross-section. The main goal of this paper is to stress that the same data set has also an excellent \emph{spin-dependent} sensitivity, which is of utmost importance in probing dark matter models. We show in particular that the constraints set by XENON100 on the spin-dependent neutron cross-section are by far the best at present, whereas the corresponding spin-dependent proton limits lag behind  other direct detection results. The effect of nuclear uncertainties on the structure functions of xenon isotopes is analysed in detail and found to lessen the robustness of the constraints, especially for spin-dependent proton couplings. Notwithstanding, the spin-dependent neutron prospects for XENON1T and DARWIN are very encouraging. We apply our constraints to well-motivated dark matter models and demonstrate that in both mass-degenerate scenarios and the minimal supersymmetric standard model the spin-dependent neutron limits can actually override the spin-independent limits. This opens the possibility of probing additional unexplored regions of the dark matter parameter space with the next generation of ton-scale direct detection experiments. 
\end{abstract}

\maketitle

\section{Introduction}
\par With hosts of on-going and forthcoming experiments, the direct search for weakly interacting massive particles (WIMPs) is today a bustling field of research and one that has witnessed an impressive progress throughout the past years. The wealth of used target materials and techniques means that we have now a rather constraining (but still inconclusive) picture of dark matter scattering off nuclei. In particular, experiments using proton- and/or neutron-odd target nuclei give useful limits on spin-dependent (SD) cross-sections, while detectors with high atomic number material chiefly probe spin-independent (SI) scattering -- the former limits lagging significantly behind the latter. At the moment, the field stands at a stalemate with several collaborations including DAMA/LIBRA \cite{DAMA2010}, CoGeNT \cite{cogentannualmod} and CRESST \cite{CRESST2011} hinting at a possible dark matter signal, a claim that has proven hard \cite{Kopp:2011yr,Frandsen:2011gi,Arina:2012dr} to reconcile with the null results of XENON10/100 \cite{XENONSD,Xenon100_2012}, CDMS \cite{cdms10} and others. This puzzling situation may change in the near future with the help of more sensitive experiments such as XENON1T (an upgrade of XENON100) and DARWIN (a consortium funded to develop ton-scale liquid xenon and liquid argon detectors).

\par Now, the results of XENON10/100 have been of particular significance in challenging the existing hints since they rule out extensive regions of the dark matter parameter space. This extreme sensitivity is made possible by a combination of a very low background environment, a large exposure and the high atomic number of xenon isotopes. Recently \cite{Xenon100_2012}, the XENON100 collaboration has released its data corresponding to 225 live days and found essentially no events above the expected background. If interpreted in terms of SI dark matter elastic scattering, this non-observation leads to the world's best cross-section limits leaving all other experiments far behind (except perhaps in the low mass regime \cite{Bottino:2008mf,Savage:2010tg} or for isospin-violating \cite{ChangIso} or magnetic inelastic \cite{Chang:2010en} dark matter candidates, among other situations). Additionally, as is well-known \cite{XENONSD,Lebedenko:2009xe}, detectors featuring xenon have also good sensitivity to SD scattering given the presence of the neutron-odd nuclei $^{129}$Xe and $^{131}$Xe. It should be noted however that the structure functions of these isotopes do suffer from significant uncertainties \cite{Ressell:1997kx,Engel:1991wq,BednyakovSpin,BednyakovFormFactor,Toivanen:2008zz,Toivanen:2009zza,Menendez:2012tm} that can affect the reconstruction of dark matter parameters \cite{Cerdeno:2012ix}. The aim of the present letter is precisely to point out that the current sensitivity level of XENON100 is already breaking records in \emph{spin-dependent} searches and to stress that upcoming ton-scale instruments are likely to deliver the strongest ever SD constraints in addition to their acclaimed SI projected limits. In Section \ref{secSD}, we compute the limits set by the latest XENON100 data \cite{Xenon100_2012} on SD cross-sections and emphasise that these are highly competitive to SD-dedicated direct detection experiments. The effect of uncertainties on the structure functions of xenon isotopes is shown explicitly, and the prospects for XENON1T and DARWIN \cite{Baudis:2012bc} are presented as well. We illustrate in Section \ref{secPPmodel} the usefulness of these SD constraints by focussing on well-motivated particle physics models before concluding in Section \ref{secConc}.

\section{Spin-dependent sensitivity}\label{secSD}

\begin{table*}
\centering
\fontsize{9}{9}\selectfont
\begin{tabular}{l|ccc|ccc}
\hline
\hline
 & \multicolumn{3}{c|}{$^{129}$Xe} & \multicolumn{3}{c}{$^{131}$Xe}  \\
 & $\langle S_p \rangle$ & $\langle S_n \rangle$ & $S_{ij}$ & $\langle S_p \rangle$ & $\langle S_n \rangle$ & $S_{ij}$\\ 
\hline
``Bonn A''	& 0.028 & 0.359 & Bonn A \cite{Ressell:1997kx}		& $-$0.009 & $-$0.227 & Bonn A \cite{Ressell:1997kx} \\
``Engel''	& 0.028 & 0.359 & Bonn A \cite{Ressell:1997kx}		& $-$0.041 & $-$0.236 & Engel \cite{Engel:1991wq} \\
``Nijmegen II''	& 0.0128 & 0.300 & Nijmegen II \cite{Ressell:1997kx}	& $-$0.012 & $-$0.217 & Nijmegen II \cite{Ressell:1997kx} \\
``Bonn CD'' & $-$0.0019 & 0.273 & Bonn CD \cite{Toivanen:2009zza} & $-$0.00069 & $-$0.125 & Bonn CD \cite{Toivanen:2009zza} \\
``Menendez+'' 	& 0.010 & 0.329 & Men\'endez et al \cite{Menendez:2012tm}	& $-$0.009 & $-$0.272 & Men\'endez et al \cite{Menendez:2012tm} \\

\hline
\end{tabular}
\caption{\fontsize{9}{9}\selectfont The spin expectation values $\langle S_{p,n} \rangle$ and structure functions $S_{ij}$ of the target nuclei $^{129}$Xe and $^{131}$Xe.}\label{tabSD}
\end{table*}

\par In order to derive direct detection constraints we follow the standard computation of dark matter elastic scattering rates in underground detectors \cite{LewinSmith} as implemented in our previous paper \cite{Garny:2012eb}, unless otherwise stated. We stick here to the so-called ``standard halo model'' \cite{LewinSmith} that features an isotropic Maxwell-Boltzmann velocity distribution of dark matter particles in our neighbourhood, and use throughout a local dark matter density $\rho_0=0.3$ GeV/cm$^3$, a local circular velocity $v_0=230$ km/s, a mean Earth velocity $v_E=244$ km/s and a local escape velocity $v_{esc}=544$ km/s. This set of values -- chiefly based on the review by Lewin \& Smith \cite{LewinSmith} and on \cite{Smith:2006ym} for $v_{esc}$ -- is prone to astrophysical uncertainties \cite{Smith:2006ym,CatenaUllio}, but has been widely used in the literature for the sake of comparison between different experimental results. In this framework the main uncertainty affecting the SD constraints from XENON100 regards the structure functions $S_{00}$, $S_{01}$, $S_{11}$ of the target nuclei $^{129}$Xe and $^{131}$Xe. We try to bracket this uncertainty and study its immediate effects by adopting the five parameterizations specified in Tab.~\ref{tabSD}: ``Bonn A'' \cite{Ressell:1997kx}, ``Engel'' \cite{Engel:1991wq} and ``Nijmegen II'' \cite{Ressell:1997kx} are rather standard nuclear models, while ``Bonn CD'' \cite{Toivanen:2009zza} is more recent and ``Menendez+'' \cite{Menendez:2012tm} refers to a very recent work.

\par Unfortunately, we find some issues in quantifying nuclear uncertainties in SD-proton scattering. Firstly, it appears that in the case of ``Bonn CD'' the combination of structure functions $S_{00}+S_{11}+S_{01}$ (which is the relevant quantity for SD-proton scattering) almost vanishes for the energies of interest and is dominated by numerical errors. This corresponds to constraints on the SD-proton cross-section weaker by a factor $\sim1000$ than those for the other parameterizations. Secondly, the ``Menendez+'' computations provide an error estimate for the structure functions $S_{01}$ and $S_{11}$ which could be translated into an uncertainty on the SD limits. A naive combination of these errors allows for a vanishing $S_{00}+S_{11}+S_{01}$ in the energy range of interest (i.e.~no SD-proton limits), but these errors are probably correlated so that a simple combination is not permissible. Since the appropriate prescription for including the errors is not available to us, we shall not consider them in the evaluation of the nuclear uncertainties. These two points indicate that the nuclear uncertainty in SD-proton scattering might be significantly larger than the canonical nuclear models \cite{Ressell:1997kx,Engel:1991wq} imply. Therefore, reliable limits on the SD-proton cross-section coming from xenon data require a better understanding of nuclear structure. None of these issues has a significant impact on the SD-neutron limits which depend on the combination $S_{00}+S_{11}-S_{01}$.

\par Now, the key motivation for the present work is the latest XENON100 data set. In Ref.~\cite{Xenon100_2012} the XENON100 collaboration reports on the observation of two nuclear recoil candidate events inside the WIMP signal region (encompassing $E_R=6.6-30.5$ keV), while the background estimate amounts to $1.0\pm0.2$. These results correspond to a data taking period of 224.6 live days and an effective exposure of 2323.7 kg.day. Applying the Feldman-Cousins procedure \cite{FeldmanCousins} with two observed events and a mean expected background of 1.0 events, we derive the 90\% confidence level (CL) upper limit $N_R\leq 4.91$ for WIMP-induced nuclear recoils. Let us stress that more sophisticated procedures (e.g.~a full profile likelihood analysis) would give somewhat stronger constraints, but our approach here is conservative and appropriate for this work. The above upper limit can be translated into separate constraints on the SD-proton and SD-neutron cross-sections $\sigma_{p,n}^{SD}$, or into a combined constraint on both cross-sections using the approach of Ref.~\cite{Savage:2004fn} (the standard approach of Ref.~\cite{Tovey:2000mm} leads to weaker constraints than presented in Figs.~\ref{fig:anap} and \ref{fig:model_SDpn} below).

\begin{figure*}[htp]
\centering
\includegraphics[width=0.49\textwidth]{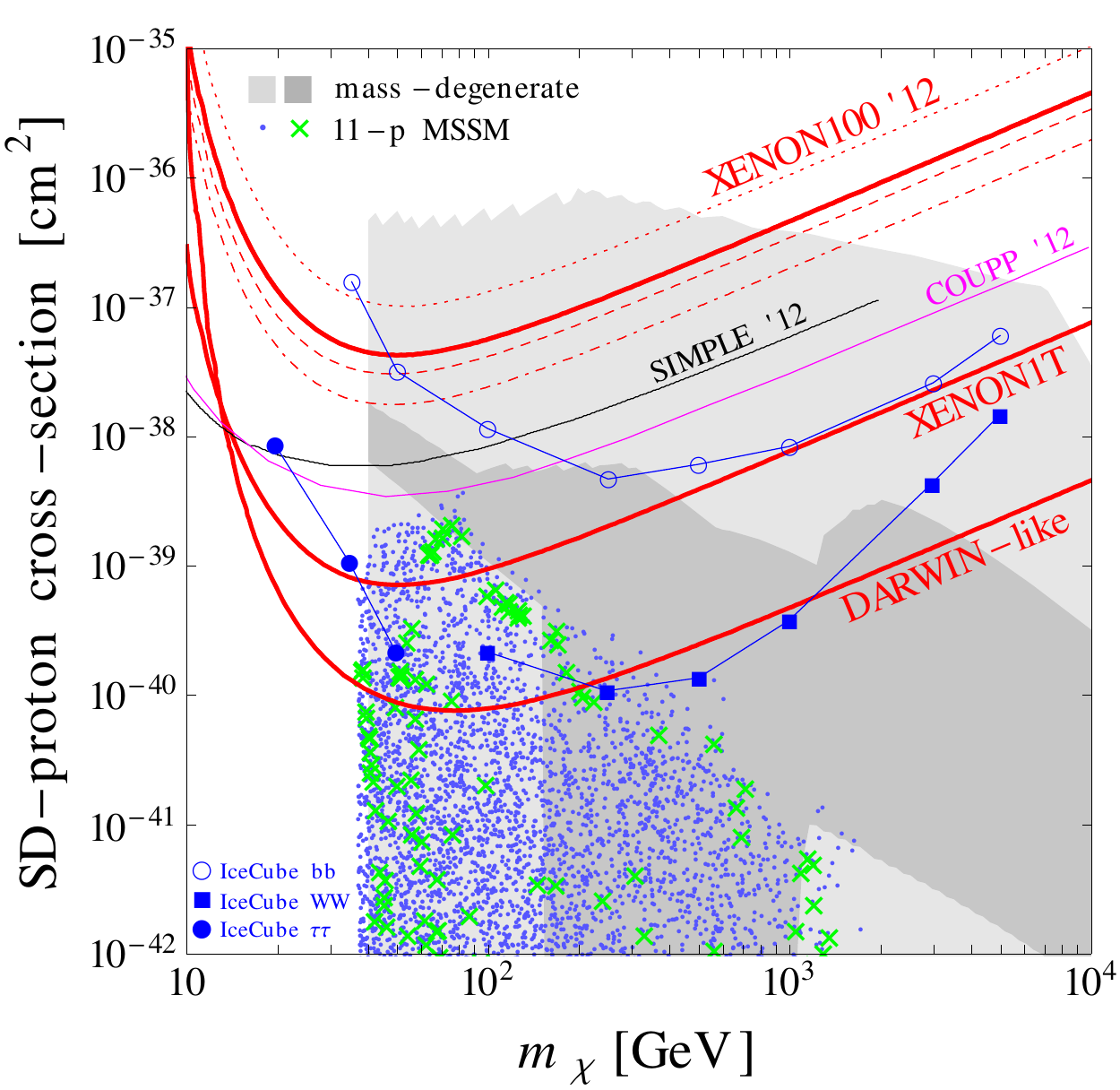}
\includegraphics[width=0.49\textwidth]{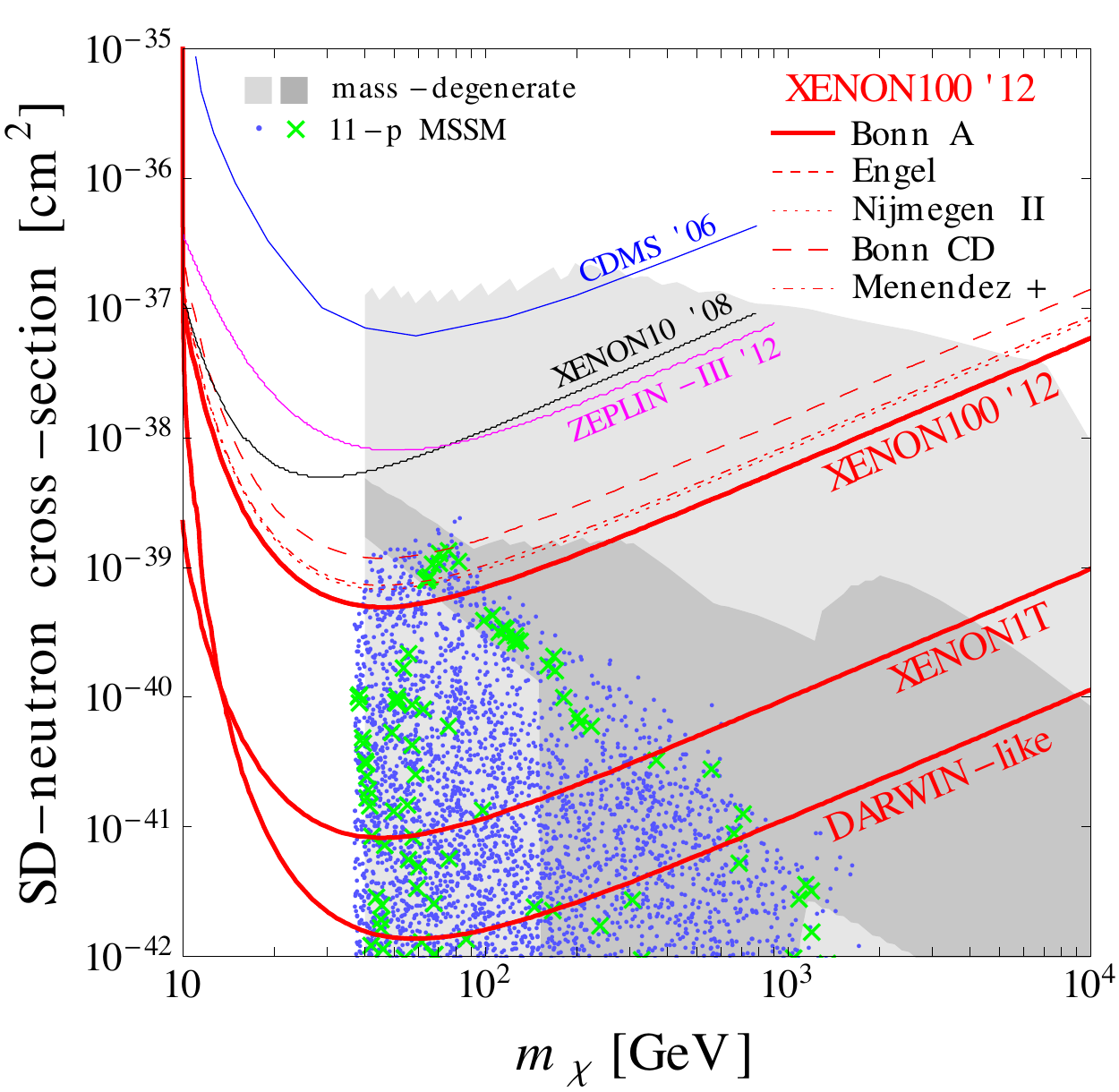}
\caption{XENON100 90\% CL exclusion limits for pure proton (left) and pure neutron (right) SD cross-sections. Shown are the results for five nuclear structure parameterizations: ``Bonn A'' \cite{Ressell:1997kx} (thick red), ``Engel'' \cite{Engel:1991wq} (short dashed red), ``Nijmegen II'' \cite{Ressell:1997kx} (dotted red), ``Bonn CD'' \cite{Toivanen:2009zza} (long dashed red) and ``Menendez+'' \cite{Menendez:2012tm} (dot-dashed red). The prospects for XENON1T  and a DARWIN-like experiment are indicated by the lower thick red curves. For a pure proton coupling (left) additional limits are available from IceCube \cite{IceCubepreliminary,Rott:2012gh} bb (empty blue circles), WW (filled blue squares), $\tau \tau$ (filled blue circles), SIMPLE '12 \cite{SIMPLE11} (black) and  COUPP '12 \cite{COUPP12} (magenta). For a pure neutron coupling (right) limits from CDMS '06 \cite{CDMS'05} (blue), XENON10 '08 \cite{XENONSD} (black) and ZEPLIN-III '12 \cite{Akimov2011} (magenta) are given for comparison. Notice that the ``Bonn A'' and ``Engel'' models lead to very similar results on $\sigma_n^{SD}$ and thus the corresponding limits are indistinguishable, whereas the ``Bonn CD'' model yields a very weak $\sigma_p^{SD}$ constraint above the plotted range. The light grey area represents the region of the mass-degenerate model as described in the text, while the dark grey region exhibits the measured relic density. Furthermore, we show a scan of the 11-parameter MSSM with the light blue dots; the green crosses indicate the points of the parameter space yielding the measured relic abundance.} 
\label{fig:SDpn}
\end{figure*}

\par We present in Fig.~\ref{fig:SDpn} the 90\% CL XENON100 upper limits on the SD-proton (left) and SD-neutron (right) cross-sections, together with the best experimental limits in the literature. The thick, short dashed, dotted, long dashed and dot-dashed red lines correspond to the five nuclear setups in Tab.~\ref{tabSD}. The nuclear uncertainties affecting XENON100 constraints can be rather large indeed. As discussed in detail above, the SD-proton limits coming from xenon target experiments are extremely sensitive to nuclear computations (note that the ``Bonn CD'' $\sigma_p^{SD}$ limit lies above the plotted range). Conservatively speaking, no solid upper limit on $\sigma_p^{SD}$ can be placed at the moment. In contrast, the nuclear nuisance on $\sigma_n^{SD}$ is relatively small and amounts to a factor $\sim 2$. Fig.~\ref{fig:SDpn} clearly shows that the latest XENON100 data are still not quite competitive in the SD-proton plane, lagging significantly behind SIMPLE and COUPP (that feature $^{19}$F), and especially IceCube. However, according to our analysis, XENON100 data beat the best published limits on the SD-neutron cross-section from XENON10 and ZEPLIN by approximately one order of magnitude, pushing the upper limit down to $\sigma_n^{SD}\sim 5\times 10^{-40}\textrm{ cm}^2$ at $m_\chi=50$ GeV.

\par Also shown in Fig.~\ref{fig:SDpn} are the prospects for XENON1T and a DARWIN-like xenon instrument, using in both cases the ``Bonn A'' setup. In the former case we simply take a 60 times better sensitivity than XENON100 (see \cite{Baudis:2012bc}), while in the latter instance we follow \cite{Pato:2010zk} and assume 2.00 ton.yr of effective xenon exposure for one background event and $E_R=10-100$ keV. This is a simplified approach which can be improved upon by the XENON1T and DARWIN collaborations themselves, but it is adequate for our purposes here. The prospects presented in Fig.~\ref{fig:SDpn} are very encouraging. On the one hand, it seems feasible to constrain SD-neutron cross-sections of $10^{-42}-10^{-41}\textrm{ cm}^2$ within the next decade. On the other hand, assuming ``Bonn A'', XENON1T will supersede current SIMPLE and COUPP SD-proton limits, while DARWIN shall easily reach the present sensitivity of neutrino telescopes. It is convenient to note that, by the time XENON1T and DARWIN will be in place, SIMPLE (whose Phase III is ongoing) and PICASSO will have improved somewhat their limits and COUPP-500 will probably probe SD-proton cross-sections as low as $10^{-42}\textrm{ cm}^2$ if no signal is observed \cite{COUPPIDM2012}. Also, current nuclear uncertainties are extremely large for SD-proton scattering on xenon. Only a better understanding of nuclear structure in the near future can eventually put XENON1T and DARWIN on the run for robust SD-proton limits.

\par Recently, rather strong bounds on the SD-nucleon cross-section inferred from monojet searches at CMS and ATLAS \cite{Rajaraman:2011wf,Chatrchyan:2012me,ATLAS:2012ky} have also been presented, reaching the level of $10^{-40}\textrm{ cm}^2$ for axial-vector contact interactions. These constraints depend however on the effective operator type and can be considerably weakened for light mediators \cite{Fox:2011pm}. Similarly, neutrino bounds rely on the assumption of equilibration between WIMP capture and annihilation at the Sun. It is fair to mention at this point that neutrino telescopes \cite{Silverwood:2012tp}, Large Hadron Collider (LHC) searches \cite{Rajaraman:2011wf}, direct detection experiments such as COUPP \cite{COUPPIDM2012} and directional detectors \cite{AlbornozVasquez:2012px} will all improve upon their present sensitivities along the next decade. 
In any case, the results in Fig.~\ref{fig:SDpn} give a strong motivation to take seriously the SD-neutron potential of ton-scale direct detection instruments. As we shall see in the next Section, the SD-neutron prospects of experiments such as XENON1T and DARWIN can easily overshadow their SI prospects in the framework of different well-motivated dark matter models.

\begin{figure}[t]
\centering
\includegraphics[width=0.49\textwidth]{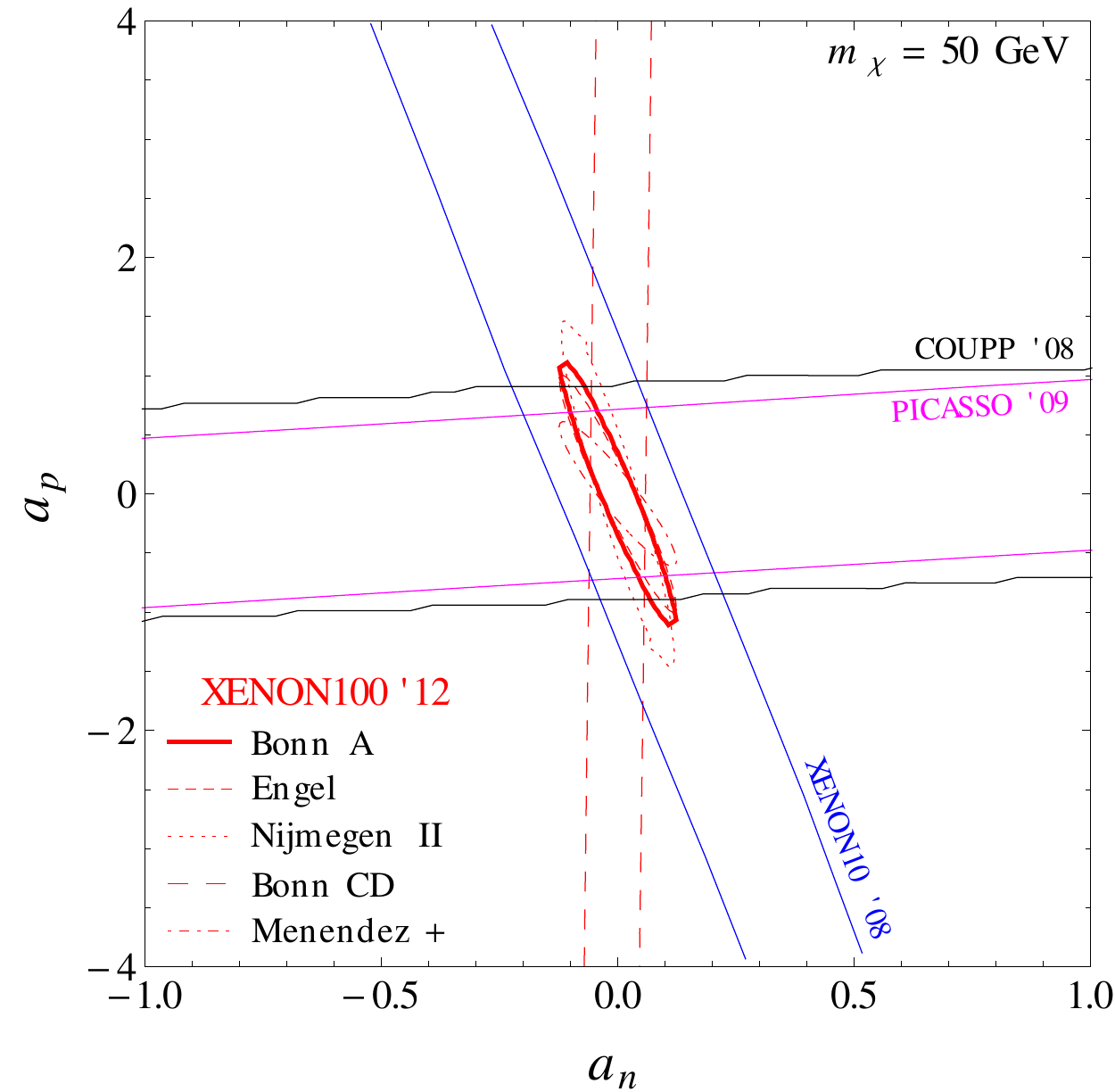}
\caption{90\% CL exclusion region in the a$_n$ - a$_p$ parameter space for a WIMP mass of 50 GeV. The line code is as in Fig.~\ref{fig:SDpn}. We also show the limits from XENON10 '08 \cite{XENONSD} (blue), COUPP '08 \cite{Behnke2008} (black) and PICASSO '09 \cite{Archambault2009} (magenta).}
\label{fig:anap}
\end{figure}

\begin{figure}[t]
\centering
\includegraphics[width=0.49\textwidth]{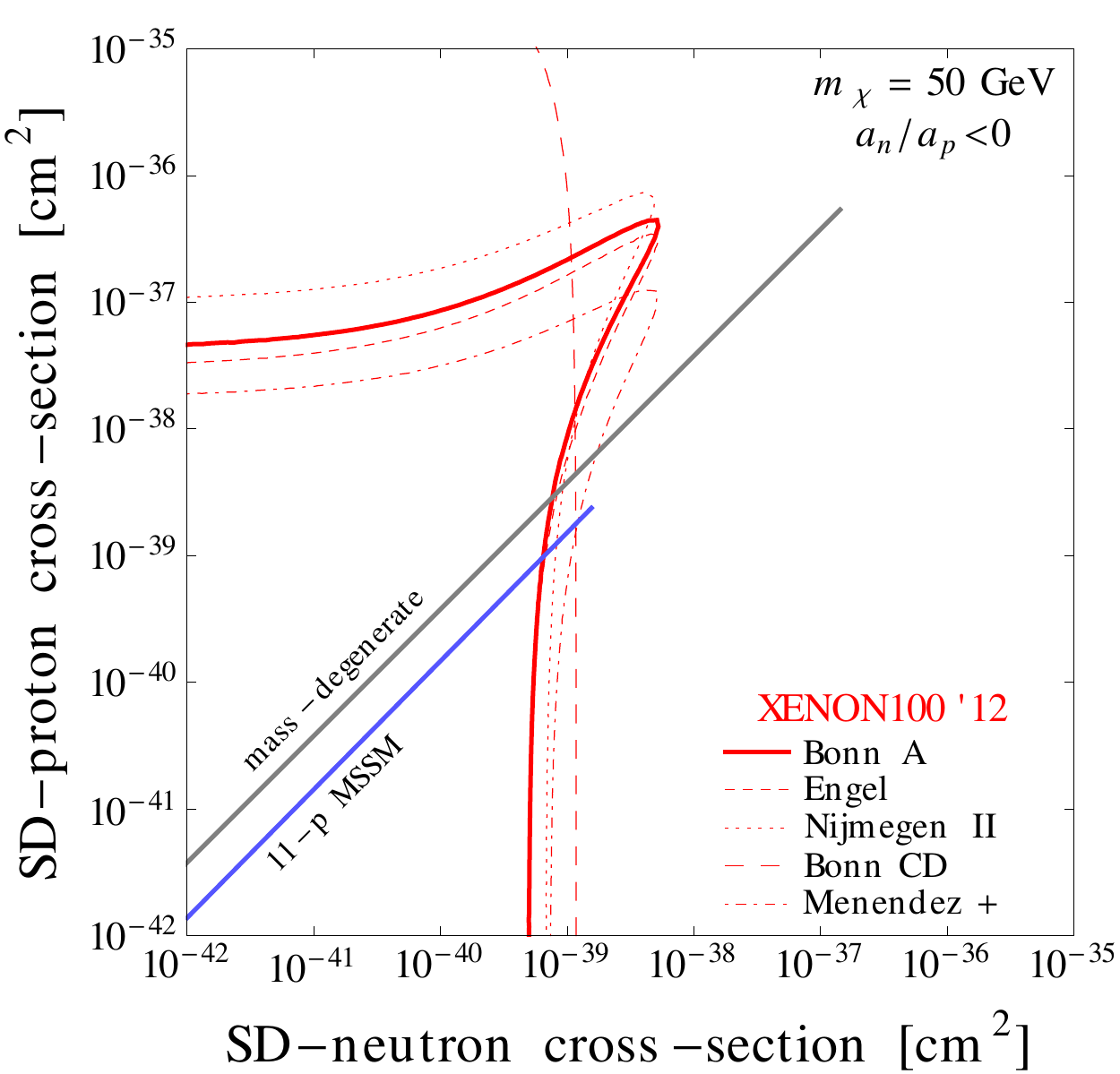}
\caption{90\% CL exclusion region in the $\sigma_n^{SD}$ - $\sigma_p^{SD}$ parameter space for a WIMP mass of 50 GeV and $a_n/a_p < 0 $. The line code is as in Fig.~\ref{fig:SDpn}. Also shown is the theoretical expectation in the mass-degenerate model (grey) and in the 11-parameter MSSM (light blue).}
\label{fig:model_SDpn}
\end{figure}

\par Fig.~\ref{fig:anap} displays the XENON100 90\% CL limits on the SD effective couplings to neutrons and protons $a_{n,p}$ for $m_\chi=50$ GeV, along with previous experimental results. The plot is particularly elucidative of the effect of nuclear uncertainties. It is nonetheless clear from Fig.~\ref{fig:anap} that XENON100 latest data shrink greatly the allowed region of parameter space, especially in combination with the results of COUPP and PICASSO. Finally, we show in Fig.~\ref{fig:model_SDpn} the combined SD constraints for $m_\chi=50$ GeV and $a_n/a_p<0$. The choice of sign of $a_n/a_p$ is motivated by the models discussed in the next Section.

\section{Specific case studies}\label{secPPmodel}

\par We now illustrate the usefulness of the SD constraints from XENON100 in the framework of specific dark matter models, starting with a simple mass-degenerate scenario \cite{Garny:2011cj,Hisano:2011um,Garny:2011ii}. This scenario consists of a Majorana dark matter particle $\chi$ that couples to the right-handed up quarks via a coloured scalar $\eta$, ${\cal L} = - f \bar\chi u_R \eta$, and is inspired by a supersymmetric model with a squark that is nearly degenerate with the neutralino. In our analysis, we treat the coupling $f$, the mass $m_\chi$ and the ratio $m_\eta/m_\chi\gtrsim 1$ as free parameters. In order to derive constraints on the parameters of this model, it is first necessary to compute the WIMP-nucleon couplings from the WIMP-quark ones -- for such we follow Ref.~\cite{Garny:2012eb} and use the nuclear parameters therein. 

\par One interesting feature of this sort of model is that the mass degeneracy strongly enhances both SD and SI cross-sections. In fact, as shown in \cite{Hisano:2011um}, even modest degeneracies lead to enhancements of several orders of magnitude. On the other hand, at sufficiently large WIMP masses and/or large splittings $m_\eta/m_\chi$ the total event rate is dominated by the SD contribution, and thus the SD constraints from XENON100 become particularly relevant. In order to show this point explicitly, we perform a simple fixed-grid scan on the parameter space $(f,m_\chi,m_\eta/m_\chi)$, taking perturbative couplings in the range $10^{-4}\leq f \leq 10$ as well as $m_\chi\geq 40$ GeV, $10^{-2} \leq m_\eta/m_\chi-1 \leq 10^2$ and $m_\eta-m_\chi>1$ GeV. LHC constraints from jets and missing energy on coloured scalars that decay into first-generation quarks and a massive neutralino \cite{ATLAS-CONF-2011-155} apply only for relatively large mass splitting $m_\eta-m_\chi \gtrsim 10^2$ GeV, while  monojet and monophoton searches yield constraints only for relatively small splittings $m_\eta-m_\chi\lesssim 10-20$ GeV \cite{Belanger:2012mk,Dreiner:2012gx}. Recently, it has been argued that a CMS razor analysis yields a constraint for simplified supersymmetric models with mass splittings between 1 and 100 GeV \cite{Dreiner:2012gx,Dreiner:2012sh}. We adapt these results by rescaling the production cross-section to fit our model and thus obtain a lower bound $m_\eta>180$ GeV. Note that LHC constraints \cite{Chatrchyan:2012me,ATLAS:2012ky} on $ \sigma_p^{SD}$ are based on contact interactions and cannot be applied directly, because of the light mediator $\eta$. In particular, while the $\chi$-nucleon scattering proceeds via the mediator in the $s$-channel, the corresponding process at colliders would be a $t$-channel exchange which is not significantly enhanced for $m_\eta\simeq m_\chi$.

\par Imposing the above collider constraints as well as the XENON100 limit on SI-only scattering, we get the light shaded region in Fig.~\ref{fig:SDpn}. The dark shading indicates the region where, in addition, the observed dark matter abundance can be explained by thermal freeze-out when taking coannihilations into account \cite{Garny:2012eb,Belanger:2010gh}. Clearly, a large portion of the parameter space escapes the usual XENON100 SI constraints but lies well above the corresponding SD-neutron upper limits. In other words, in a large portion of the parameter space the scattering rate in a xenon target experiment is dominated by the SD-neutron contribution. This stresses the importance of analysing XENON100 results in light of SD -- and not only SI -- scattering for mass-degenerate scenarios. It is also noteworthy that a chunk of the thermal region is already being excluded by XENON100 SD-neutron constraints and that XENON1T and DARWIN will be able to probe a large portion of this region. The scan is also shown as a grey line in Fig.~\ref{fig:model_SDpn} for $m_\chi=50$ GeV. Note that it lies along a straight line because in these models $a_n/a_p=\Delta u^{(n)}/\Delta u^{(p)}\simeq-0.52$ (here $\Delta u^{(n,p)}$ are the up-quark spin contents of the neutron and proton, respectively, see e.g.~\cite{Ellis:2008hf}), or $\sigma_{n}^{SD}/\sigma_{p}^{SD} \simeq 0.27$.

\par Let us now briefly comment on the prospects for supersymmetric neutralino dark matter. Within the constrained minimal supersymmetric standard model (CMSSM), when taking LHC constraints into account and assuming a Higgs mass $m_h\simeq 125$ GeV, most regions of the parameter space where the dark matter abundance can be explained by thermally produced neutralinos are highly constrained \cite{Baer:2012uya} (see also \cite{Buchmueller:2011ab,Strege:2011pk,Fowlie:2012im}). In addition, when applying constraints on SI scattering from XENON100, the SD cross-section gets pushed to values below $10^{-42}$ cm$^2$ \cite{Strege:2011pk}, which lie much below the future sensitivity of XENON1T or DARWIN (see Fig.~\ref{fig:SDpn}).

\begin{table*}
\centering
\fontsize{9}{9}\selectfont
\begin{tabular}{l|ccccccccccc}
\hline
\hline
Parameter & $M_1$ & $M_2$ & $|\mu|$ & $m_{\tilde q^L_{1,2}}$ & $m_{\tilde q^R_{1,2}}$ & $m_{\tilde q^L_{3}}$ & $m_{\tilde q^R_{3}}$ & $m_{\tilde\ell}$ & $\tan\beta(M_Z)$ & $m_A$ & $A_t$ \\
Min [GeV] & 10    & 80   & 80    & 1000 & 1000  & 500  & 500   & 80   & 5  & 500  & -4000\\
Max [GeV] & 2000  & 2000 & 2000  & 3000 & 3000  & 3000 & 3000  & 3000 & 25 & 2000 & 4000\\

\hline
\end{tabular}
\caption{\fontsize{9}{9}\selectfont
Parameter ranges for the MSSM scan. The gluino mass parameter is fixed to $M_3=2$ TeV, $A_{u,d,b}=0$ and $m_{\tilde\ell}\equiv m_{\tilde{\ell,i}}^{L,R}$. All parameters except $\tan\beta$ and the pole mass $m_A$ are given at $Q=1$ TeV.
}\label{tabMSSM}
\end{table*}

\par In order to check in how far this conclusion can be relaxed when lifting some of the severe assumptions underlying the CMSSM, we have performed an 11-parameter random scan of the MSSM using DarkSUSY \cite{Gondolo:2004sc} with parameters shown in Tab.~\ref{tabMSSM}. Apart from accelerator constraints as implemented in DarkSUSY 5.0.5, including $b\to s\gamma$ and the $\rho$-parameter, we require a Higgs mass in the range $m_h=125.5\pm1.5$ GeV. Since a dedicated analysis of LHC constraints within the enlarged MSSM parameter space is beyond the scope of this work, we instead impose conservative bounds $m_{\tilde q_{1,2}}>1.5$ TeV and $m_{\tilde q_3}>500$ GeV that are allowed within simplified models \cite{atlas:2012rz,atlas:2012ar}. In addition, for each point we impose the XENON100 bound on SI-only scattering and conservatively require the annihilation cross-section to be below the most stringent Fermi-LAT limits from dwarf galaxies \cite{Ackermann:2011wa}. We have checked that within the range of parameters, especially for $m_A$ and $\tan\beta$, recent constraints on BR$(B_s\to\mu^+\mu^-)$ \cite{Aaij:2012ac} are easily satisfied. However, allowing for larger values of $\tan\beta$ or a slightly wider Higgs mass range would not change our results significantly.

\par The range of the SD cross-section that is compatible with all constraints is indicated by the light blue dots in Fig.~\ref{fig:SDpn} and by the light blue line in Fig.~\ref{fig:model_SDpn}. Cross-sections as large as $\sigma^{SD}_n\simeq 10^{-39}$ cm$^2$ are reached for neutralino masses around $100$ GeV -- this is in line with the findings of Ref.~\cite{Bertone:2007xj}. The green crosses in Fig.~\ref{fig:SDpn} correspond to configurations yielding the measured relic abundance. The neutralino is in this case a mixed bino-higgsino state, characterised by a small $\mu$-term that is not much larger than $M_1$. Neutralino annihilation occurs mainly via $Z$ and chargino exchange, while all processes involving squark exchange are suppressed due to the large squark masses. Note that the corresponding region within the CMSSM is disfavoured by LHC and Higgs mass constraints \cite{Baer:2012uya}.

\par For the mixed bino-higgsino scenario, SI scattering is mediated mainly by Higgs exchange while SD scattering is mediated mainly by $Z$ exchange, in the limit of heavy squarks \cite{Ellis:2008hf}. Note that $a_n/a_p\simeq -0.8$ for $Z$ exchange, in good agreement with Fig.~\ref{fig:model_SDpn}. In order to understand the relation between SD and SI scattering in this case, consider the effective Lagrangian
\begin{equation}
 {\cal L} = \alpha_{2i}\bar\chi\gamma^\mu\gamma^5\chi\bar q_i \gamma_\mu\gamma_5 q_i + \alpha_{3i}\bar\chi\chi\bar q_i q_i\,,
\end{equation}
where the first (second) term describes SD (SI) scattering, and $i=u,d$ (in our numerical analysis we also take twist-2 and
loop-induced couplings into account). In the decoupling/heavy squark limit, the coefficients for a neutralino $\chi = Z_{\chi 1} \tilde B + Z_{\chi 2} \tilde W + Z_{\chi 3} \tilde H_u + Z_{\chi 4} \tilde H_d$ are given by (we use the
notation of \cite{Ellis:2008hf})
\begin{eqnarray}
 \alpha_{2i} &=& - \frac{g^2 T_{3i}}{8 M_Z^2 c_W^2} \left( |Z_{\chi 3}|^2-|Z_{\chi 4}|^2 \right) \\
             &\simeq& - \frac{g^2 T_{3i} t_W^2}{8(t_\beta^2+1)(\mu^2-M_1^2)^2} \left(t_\beta^2(\mu^2-M_1^2) - 2\mu M_1\right) \,,\nonumber 
             \end{eqnarray}
             \begin{eqnarray}
 \alpha_{3i} &=& - \frac{g^2 m_{qi}}{4 M_W m_h^2} \mbox{Re}\left[ ( s_\beta Z_{\chi 4} - c_\beta Z_{\chi 3} ) ( Z_{\chi 2} - t_W Z_{\chi 1} ) \right] \nonumber \\
             &\simeq& - \frac{g^2 m_{qi} t_W^2 t_\beta}{4 m_h^2 (t_\beta^2+1) (\mu^2-M_1^2)} \left( 2\mu + t_\beta M_1 \right) \,.
\end{eqnarray}
The approximate expressions apply for $M_1 \lesssim |\mu| \ll M_2$ and large $t_\beta\equiv \tan\beta$. Typically, both contributions are strongly correlated. Nevertheless, we would like to point out that in the 11-parameter scan we find a non-negligible portion of the parameter space where the SI cross-section is suppressed with respect to the SD one. Closer inspection shows that this region corresponds to $\mu<0$ and $|\mu|/M_1\gtrsim \mathcal{O}(1)$. This behaviour is consistent with the approximate analytic expressions (see also \cite{Falk:1998xj,ArkaniHamed:2006mb}). In particular, there exists a region in parameter space where $\sigma^{SI}_{p,n}\lesssim 10^{-46}$cm$^2$ is below the SI-sensitivity of XENON1T, while the corresponding SD scattering cross-section on neutrons $\sigma^{SD}_n\sim 10^{-40}$cm$^2$ lies within reach of
XENON1T. At colliders, an associated signal is a light chargino. Concerning monojet searches one naively expects that the bounds inferred when assuming contact interactions get weakened by a factor of the order $M_Z^2/s$, where $s$ is the partonic centre of mass energy \cite{Fox:2011pm}.

\section{Conclusion}\label{secConc}
\par Direct dark matter detection is now entering a new sensitivity phase with different experiments -- most prominently, XENON100 -- pushing down cross-section upper limits to extremely low values. In this framework, it is important not to focus solely on SI scattering, but to explore as well the SD potential of each data set. In fact, there are several dark matter models for which the SD contribution to the total event rate completely overshadows the SI one. As illustrated in this work, examples of such models are mass-degenerate scenarios and some parts of the parameter space of the MSSM. On that note, we have derived here the SD constraints set by the latest XENON100 data. While in SD-proton the constraints are prone to very significant nuclear uncertainties, in SD-neutron XENON100 overrides the best published limits by roughly one order of magnitude independently of nuclear nuisances. Along the same lines, we find exciting prospects for XENON1T and DARWIN in constraining SD-neutron couplings. Interestingly, for a large class of dark matter models, the next generation of ton-scale direct detection instruments would detect first SD scattering and only then the SI counterpart.

\vspace{0.5cm}
{\it Acknowledgements:} The authors would like to thank David G.~Cerde\~no and Nicolao Fornengo for useful comments. This work has been partially supported by the DFG cluster of excellence ``Origin and Structure of the Universe'' and by the DFG Collaborative Research Center 676 ``Particles, Strings and the Early Universe''. S.V.~acknowledges support from the DFG Graduiertenkolleg ``Particle Physics at the Energy Frontier of New Phenomena''. This work makes use of DarkSUSY \cite{Gondolo:2004sc}, SOFTSUSY \cite{Allanach:2001kg}, SuperIso \cite{Mahmoudi:2007vz,Mahmoudi:2008tp}, micrOMEGAs \cite{Belanger:2010gh} and Mathematica.


\bibliographystyle{apsrev}
\bibliography{XE100SD}

\end{document}